\pdfoutput=1
\documentclass[11pt,twoside,a4paper,cmspaper,final,collab]{cms-tdr}
\def\svnVersion{b018d2d}\def\svnDate{2022/08/08}\def\cmsCernNoTag{CERN-EP-2021-223}\def\cmsCernDate{\today}\def\cmsMessage{Published in the Journal of High Energy Physics as \href{http://dx.doi.org/10.1007/JHEP04(2022)048}{\doi{10.1007/JHEP04(2022)048}.}}
\begin{document}\cmsNoteHeader{B2G-20-010}

\newlength{\cmsFigWidth}
\ifthenelse{\boolean{cms@external}}{\setlength{\cmsFigWidth}{0.45\textwidth}}{\setlength{\cmsFigWidth}{0.49\textwidth}}
\ifthenelse{\boolean{cms@external}}{\providecommand{\cmsLeft}{upper\xspace}}{\providecommand{\cmsLeft}{left\xspace}}
\ifthenelse{\boolean{cms@external}}{\providecommand{\cmsRight}{lower\xspace}}{\providecommand{\cmsRight}{right\xspace}}

\newcommand{\bstar}{\ensuremath{\PQb^\ast}\xspace}
\newcommand{\ST}{\ensuremath{S_{\mathrm{T}}}\xspace}
\newcommand{\MtW}{\ensuremath{M_{\PQt\PW}}\xspace}
\newcommand{\WZjets}{\ensuremath{\PW{}\!/\PZ{}+\text{jets}}\xspace}
\newcommand{\ljets}{\ensuremath{\Pell\text{+jets}}\xspace}
\newcommand{\PBt}{\PB{}+\PQt}
\newcommand{\PBb}{\PB{}+\PQb}
\newcommand\Tstrut{\rule{0pt}{2.2ex}}
\newcommand\Bstrut{\rule[-1.5ex]{0pt}{0pt}}
\newcommand{\Reff}{\ensuremath{R_{\text{eff}}}\xspace}
\newcommand{\Xsquare}{\ensuremath{X^2}\xspace}

\cmsNoteHeader{B2G-20-010}

\title{Search for a heavy resonance decaying into a top quark and a \texorpdfstring{\PW}{W} boson in the lepton+jets final state at
\texorpdfstring{$\sqrt{s} = 13\TeV$}{sqrt(s) = 13 TeV}}

\date{\today}

\abstract{
A search for a heavy resonance decaying into a top quark and a \PW boson in proton-proton collisions at $\sqrt{s}=13\TeV$ is presented. The data analyzed were recorded with the CMS detector at the LHC and correspond to an integrated luminosity of 138\fbinv. The top quark is reconstructed as a single jet and the \PW boson, from its decay into an electron or muon and the corresponding neutrino. A top quark tagging technique based on jet clustering with a variable distance parameter and simultaneous jet grooming is used to identify jets from the collimated top quark decay. The results are interpreted in the context of two benchmark models, where the heavy resonance is either an excited bottom quark \bstar or a vector-like quark \PB. A statistical combination with an earlier search by the CMS Collaboration in the all-hadronic final state is performed to place upper cross section limits on these two models. The new analysis extends the lower range of resonance mass probed from 1.4 down to 0.7\TeV. For left-handed, right-handed, and vector-like couplings, \bstar masses up to 3.0, 3.0, and 3.2\TeV are excluded at 95\% confidence level, respectively. The observed upper limits represent the most stringent constraints on the \bstar model to date.
}

\hypersetup{
pdfauthor={CMS Collaboration},
pdftitle={Search for a heavy resonance decaying into a top quark and a W boson in the lepton+jets final state at sqrt(s) = 13 TeV},
pdfsubject={CMS},
pdfkeywords={CMS, physics, excited bottom quark, resonance}}

\maketitle

\section{Introduction}

The remarkable success of the standard model (SM) of particle physics is built upon extensive
experimental tests and verifications. However, there are indications that extensions of the SM are
needed to explain observed phenomena.
Many possibilities for physics beyond the SM have been proposed,
including a scenario of compositeness, where excited states of quarks are predicted to exist~\cite{Harari:1982xy}.
These states can have masses of the order of 1\TeV and can be probed in
high-energy proton-proton ($\Pp\Pp$) collisions~\cite{PhysRevD.42.815}.
Evidence for such new physics could be signaled by the observation of a heavy resonance that decays
into a top quark (\PQt) and a \PW boson~\cite{Tait:2000sh,Nutter:2012an}.
In this paper, we present a search for such a resonance using
$\Pp\Pp$ collision data at $\sqrt{s}=13\TeV$
collected by the CMS experiment~\cite{Chatrchyan:2008zzk} at the CERN LHC,
corresponding to an integrated luminosity of 138\fbinv.

A model where a heavy resonance takes the form of an excited bottom quark (\bstar) is used as a benchmark scenario.
In $\Pp\Pp$ collisions, a \bstar would be singly produced via the strong interaction, described by
the following effective Lagrangian
\begin{linenomath}
\begin{equation}
\mathcal{L}_1 = \frac{g_\mathrm{s}}{2\Lambda}G_{\mu\nu} \PAQb \sigma^{\mu\nu}\left(\kappa^{\PQb}_{\mathrm{L}}P_{\mathrm{L}}+
\kappa^{\PQb}_{\mathrm{R}}P_{\mathrm{R}}\right)\bstar + \text{h.c.},
\end{equation}
\end{linenomath}
where $g_\mathrm{s}$ is the strong coupling constant, $G_{\mu\nu}$ is the field strength tensor
of the gluon, \PQb is the bottom quark field, \bstar is the excited \PQb quark field, $\sigma^{\mu\nu}$ is the Pauli spin matrix,
and $\Lambda$ is the scale of compositeness,
chosen to be the \bstar mass~\cite{PhysRevD.42.815}.
The left- and right-handed chiral projection operators are $P_\mathrm{L}$ and $P_\mathrm{R}$,
with respective relative coupling strengths $\kappa^{\PQb}_{\mathrm{L}}$ and $\kappa^{\PQb}_{\mathrm{R}}$.

\begin{figure}[tb]
\centering
\includegraphics[width=\cmsFigWidth]{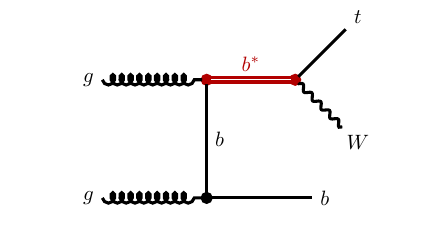} 
\includegraphics[width=\cmsFigWidth]{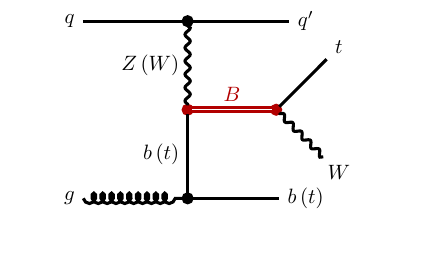}
\caption{Example Feynman diagrams at LO of the production and decay of a \bstar (\cmsLeft) 
and a \PB quark (\cmsRight), drawn with the TikZ-Feynman package~\cite{Ellis:2016jkw}. 
\label{fig:feyn}}
\end{figure}

The possible \bstar decay modes include $\PQb\cPg$, $\PQb\PZ$, $\PQb\PH$ and $\PQt\PW$.
The decay to $\PQt\PW$ is predicted to be dominant for \bstar masses $m_{\bstar}>700\GeV$,
approaching a branching fraction of 40\%~\cite{Nutter:2012an}.
This decay can be described by the following effective Lagrangian
\begin{linenomath}
\begin{equation}
\mathcal{L}_2 = \frac{g_2}{\sqrt{2}} \PW^+_\mu \cPaqt \gamma^\mu \left(g_{\mathrm{L}} P_{\mathrm{L}} +
g_{\mathrm{R}} P_{\mathrm{R}}\right) \bstar + \text{h.c.},
\end{equation}
\end{linenomath}
where $g_2$ is the $SU(2)_L$ weak coupling constant, $\gamma^\mu$ are the gamma matrices,
and $g_{\mathrm{L}}$ and $g_{\mathrm{R}}$ are the relative coupling strengths
of the \PW boson field $\PW^+_\mu$ to the left- and right-handed \bstar chirality states, respectively.
Three choices of coupling parameters are considered: purely left-handed (LH)
($g_{\mathrm{L}}=1$, $\kappa_{\mathrm{L}}^{\PQb}=1$, $g_{\mathrm{R}}=0$, $\kappa_{\mathrm{R}}^{\PQb}=0$),
purely right-handed (RH)
($g_{\mathrm{L}}=0$, $\kappa_{\mathrm{L}}^{\PQb}=0$, $g_{\mathrm{R}}=1$, $\kappa_{\mathrm{R}}^{\PQb}=1$)
and vector-like (VL)
($g_{\mathrm{L}}=1$, $\kappa_{\mathrm{L}}^{\PQb}=1$, $g_{\mathrm{R}}=1$, $\kappa_{\mathrm{R}}^{\PQb}=1$). 
With this choice of couplings, the relative width of the resonance is about 4\% in the probed \bstar mass range. 
A leading order (LO) Feynman diagram of the \bstar production and decay at the LHC is shown in 
Fig.~\ref{fig:feyn} (\cmsLeft). 

Searches for a \bstar in the $\PQt\PW$ decay channel have been performed at $\sqrt{s}=7$ and 8\TeV by
the ATLAS~\cite{Aad:2013rna,Aad:2015voa} and CMS~\cite{Khachatryan:2015mta} Collaborations.
In these analyses, the highest mass limits for LH, RH and VL couplings
were obtained by the CMS Collaboration with values of 1.4, 1.4 and 1.5\TeV~\cite{Khachatryan:2015mta} at 95\% confidence level (\CL), respectively.
A recent analysis~\cite{Sirunyan:2021fkj} by the CMS Collaboration in the all-hadronic final state used the same
$\Pp\Pp$ collision data as the present paper.
The analysis considered \bstar masses above 1.4\TeV and improved the limits to 2.6, 2.8 and 3.1\TeV for LH, RH and VL
couplings, respectively. Masses below 1.4\TeV were inaccessible to that analysis because of high trigger thresholds,
where the presence of highly energetic jets was required to record events.

The analysis presented here is performed in the \ljets final state, where $\Pell$ denotes an electron or a muon.
The analysis targets the $\PW\to\Pell\PGn$ and $\PQt\to\PW\PQb\to\cPq\cPaq^\prime\PQb$ decay, where the \PQt quark decay
is reconstructed using a single jet with adaptive angular size, obtained with the Heavy Object Tagger with Variable $R$ (HOTVR)~\cite{Lapsien:2016zor} algorithm.
The probed final state consists of one lepton, missing transverse
momentum (\ptmiss), and one jet with high transverse momentum (\pt), identified by its substructure to originate
from collimated \PQt quark decay products~\cite{Larkoski_2020,Kogler_2019}.
The presence of a lepton in the final state allows the use of lepton triggers with lower \pt thresholds than jet triggers,
thus extending the range of the analysis down to \bstar masses of 0.7\TeV. In order to achieve a stable selection efficiency
over a large range of probed \bstar masses, the HOTVR
algorithm is employed for the identification of collimated \PQt quark decays.
The \PQt jet, lepton and \ptmiss are used to reconstruct the invariant mass of the \PQt{}\PW system, \MtW.
The spectrum of \MtW is used to search for the heavy resonance,
using a binned maximum likelihood fit to data in both a signal and a control region
simultaneously with the distributions of the all-hadronic search, mentioned above.

Finally, the results are interpreted for evidence of the production of a vector-like quark, \PB~\cite{Aguilar-Saavedra:2013qpa,DeSimone:2012fs},
decaying into \PQt{}\PW, which has a similar signature to the \bstar quark decay described above.
The mixing parameter $V^{\mathrm{L}}_{\PQt\PB}$ in the considered model, defined in Ref.~\cite{Aguilar-Saavedra:2013qpa}, is set to unity,
resulting in a resonance with a relative width of less than 5\% in the probed \PB quark mass range and a branching fraction to \PQt{}\PW of approximately 50\%.
In contrast to the \bstar model, the vector-like \PB quark is produced via an electroweak interaction in association with either a \PQt quark
(\PBt) or a \PQb quark (\PBb), shown in Fig.~\ref{fig:feyn} (\cmsRight).
Besides the third generation quarks produced together with the heavy resonance, the electroweak
interaction results in an additional jet at LO, which is not present in the
\bstar model at this order. We consider both \PB production modes, but the search is not optimized for these more complex final states.

Tabulated results are provided in the HEPData record for this analysis~\cite{hepdata}.

\section{Data and simulated samples}

This analysis uses $\Pp\Pp$ collision data at $\sqrt{s}=13\TeV$
recorded by the CMS detector in the years 2016, 2017 and 2018.
The data were recorded with electron and muon triggers~\cite{Khachatryan:2016bia}.
For the electron trigger, an isolated electron candidate with $\pt > 27$, 35 and 32\GeV was required in the years 2016, 2017 and 2018, respectively.
To recover trigger inefficiencies at high electron \pt, events are accepted if recorded with a trigger requiring
a photon candidate with an energy $E > 175\GeV$ in the year 2016,
and $E > 200\GeV$ in the years 2017 and 2018.
For the muon trigger~\cite{CMS:2021yvr}, an isolated muon candidate with $\pt > 24$ (27)\GeV was required in 2016 and
2018 (2017). Varying trigger thresholds reflect the changing experimental conditions between the years.
The data set corresponds to an integrated luminosity of 138\fbinv.
Data and simulation are categorized by year, and dedicated corrections are applied before combining the distributions
from all three years to derive the final result.

The SM production of top quark-antiquark pairs (\ttbar) in the \ljets final state constitutes
the main background for this search. It is simulated at next-to-leading order (NLO) with the
\POWHEG~v2~\cite{Frixione:2007vw,Nason:2004rx,Alioli:2010xd,Frixione:2007nw,Re:2010bp} matrix element generator.
Single \PQt quark production in association with a \PW boson has the same signature as the signal,
making it an irreducible background. Single top quark production in the $s$ and $t$ channels is considered as well,
but its contribution is small. The \PQt{}\PW and $t$-channel events are produced at NLO with
\POWHEG~v2, while $s$-channel events are produced at LO with \MGvATNLO~\cite{Alwall:2014hca,Alwall:2007fs}
in version 2.2.2 for 2016 and version 2.4.2 for 2017 and 2018.
The cross section for the \ttbar background is adjusted to a prediction at
next-to-next-to-leading order (NNLO) precision in perturbative quantum chromodynamics (QCD), including resummation of
next-to-next-to-leading-logarithmic soft gluon terms, obtained with \TOPpp~2.0~\cite{Czakon:2011xx}.
The cross section of single \PQt quark production in association with a \PW is adjusted to NNLO approximate calculations
taken from Refs.~\cite{Kidonakis:2010ux,Kidonakis:2013zqa}.
The $s$- and $t$-channel cross sections are adjusted to predictions at NLO precision
obtained with \textsc{hathor}~v2.1~\cite{Aliev:2010zk}.

Because of a difference between data and simulation in the top quark \pt spectrum in \ttbar
production~\cite{Khachatryan:2016mnb,Sirunyan:2018ucr}, the top quark \pt in the
\ttbar simulation is corrected using the procedure described in Section 3 of Ref.~\cite{Sirunyan:2021fkj}.
A weight is assigned to each event, proportional to $\exp(-\beta \pt)$, which results in a softer
simulated \pt spectrum than predicted by \POWHEG. The free parameter $\beta$ is initialized to
$0.5\TeV^{-1}$ and later determined in a fit to data, as described below in Sec.~\ref{sec:results}.

Simulated background events from the production of electroweak vector bosons
in association with jets (\WZjets) and diboson production
are used in the data-driven background estimation described in Section~\ref{sec:background_estimation}.
The \WZjets samples are simulated with \MGvATNLO at NLO precision. Diboson events are produced at LO
using the \PYTHIA event generator~\cite{Sjostrand:2014zea} in version 8.212 for 2016 and version 8.230 for 2017 and 2018.

The simulations of \bstar and vector-like \PB production and decay are performed at LO using the \MGvATNLO generator.
The five-flavor scheme is used in the simulation of the initial state. Masses in the range of 0.7 to
4.0\TeV are used with LH and RH \bstar couplings. The VL \bstar samples are obtained
as the sum of the respective LH and RH samples. The \PB signal samples are simulated using 2016 conditions and
then scaled to the luminosity of the full data set. We estimate the selection efficiencies
in the years 2017 and 2018 by calculating the differences in selection efficiencies between 2016 and 2017/2018
using the \bstar simulations.
Separate signal samples for the \PBt and \PBb production modes are simulated
for \PB masses between 0.7 and 1.8\TeV.

The parton shower and hadronization are simulated using \PYTHIA~8.212 with the CUE\-TP8\-M1~\cite{Khachatryan:2015pea}
underlying event tune for the 2016 simulation of samples other than \ttbar, while
the CUE\-TP8\-M2T4~\cite{CMS:2016kle} tune is used for the simulation of \ttbar.
For the 2017 and 2018 simulations, \PYTHIA~8.230 is used with the CP5~\cite{Sirunyan:2019dfx} tune.
The NNPDF3.0~\cite{Ball_2015} parton distribution functions (PDFs) are used for
2016 simulation and the NNPDF3.1~\cite{Ball_2017} PDFs are used for 2017 and 2018 simulations.
Additional inelastic $\Pp\Pp$ collision events are simulated using \PYTHIA and superimposed on simulated
events to model the effect of additional $\Pp\Pp$ collisions within the same or adjacent bunch crossings (pileup).
We use a total inelastic cross section of 69.2\unit{mb}~\cite{Sirunyan:2018nqx} to estimate the expected
number of $\Pp\Pp$ interactions per bunch crossing and correct the simulation to match the corresponding distribution to that observed in data.
The CMS detector simulation is performed with \GEANTfour~\cite{Agostinelli:2002hh}.

\section{The CMS detector and event reconstruction}

The central feature of the CMS apparatus is a superconducting solenoid of 6\unit{m} internal diameter,
providing a magnetic field of 3.8\unit{T}. Within the solenoid volume are silicon pixel and strip tracking detectors,
a lead tungstate crystal electromagnetic calorimeter (ECAL), and a brass and scintillator hadron calorimeter (HCAL),
each composed of a barrel and two endcap sections. Forward calorimeters extend the pseudorapidity ($\eta$)
coverage provided by the barrel and endcap detectors. Muons are detected in gas-ionization chambers embedded
in the steel flux-return yoke outside the solenoid.
A more detailed description of the CMS detector, together with a definition of the coordinate system
used and the relevant kinematic variables, can be found in Ref.~\cite{Chatrchyan:2008zzk}.
Between the 2016 and 2017 data taking runs, the CMS pixel detector was upgraded. Details about the changes
can be found in Ref.~\cite{CMSTrackerGroup:2020edz}.

Events of interest are selected using a two-tiered trigger system. The first
level (L1), composed of custom hardware processors, uses information from the
calorimeters and muon detectors to select events at a rate of around
100\unit{kHz} within a fixed latency of about 4\mus~\cite{Sirunyan:2020zal}. The
second level, known as the high-level trigger, consists of a farm of
processors running a version of the full event reconstruction software optimized
for fast processing, and reduces the event rate to around 1\unit{kHz} before
data storage~\cite{Khachatryan:2016bia}.

A particle-flow (PF) algorithm~\cite{Sirunyan:2017ulk} aims to reconstruct and identify
each individual particle in an event, using an optimized combination of information from the
various elements of the CMS detector.
The energy of photons is obtained from the ECAL measurement. The energy of electrons is determined
from a combination of the electron momentum at the primary interaction vertex as determined by the tracking detectors,
the energy of the corresponding ECAL cluster, and the energy sum of all bremsstrahlung photons
spatially compatible with originating from the electron track~\cite{Sirunyan:2020ycc}. The energy of muons is obtained from the
curvature of the corresponding track~\cite{Sirunyan:2018fpa}. The energy of charged hadrons is determined from a combination of
their momentum measured in the tracking detectors and the matching ECAL and HCAL energy deposits, corrected for the
response function of the calorimeters to hadronic showers. Finally, the energy of neutral hadrons is
obtained from the corresponding corrected ECAL and HCAL energies.
The candidate vertex with the largest sum of the square of the transverse momenta $\pt^2$ of the
physics objects is taken to be the primary $\Pp\Pp$ interaction vertex.
The physics objects are the jets, clustered using the jet finding algorithm~\cite{Cacciari:2008gp,Cacciari:2011ma} with the tracks assigned
to candidate vertices as inputs, and the associated missing transverse momentum, taken as the negative vector sum of the \pt of those jets.
More details are given in Section~9.4.1 of Ref.~\cite{CMS-TDR-15-02}.

Electrons and muons are required to fulfill $\abs{\eta}<2.4$ and $\pt>30\GeV$.
We require tight quality criteria with small misidentification probabilities of
about 1\% for electrons~\cite{Sirunyan:2020ycc} and 0.1\% for muons~\cite{Sirunyan:2018fpa}.
In addition, electrons and muons are required to be isolated, where the relative isolation is
measured by the \pt sum of all PF particles in a cone around the lepton relative to the lepton \pt~\cite{Sirunyan:2018fpa,Khachatryan:2015hwa}.
The cone is defined by a distance in $\Delta R = \sqrt{\smash[b]{(\Delta \eta)^2 + (\Delta \phi)^2}}$
of 0.3 for electrons and 0.4 for muons, where $\phi$ is the azimuthal angle.
The isolation must be less than $0.0445 + 0.963\GeV/\ET$ for electrons,
with \ET the transverse energy of the electron, and less than 0.15 for muons.
The lepton isolation is corrected for the contribution of neutral hadrons from pileup.
The identification, isolation, and trigger efficiencies are measured in dedicated analyses and
are adjusted in simulation to match those in data.

Jets are reconstructed from PF candidates, using the
anti-\kt~\cite{Cacciari:2008gp} or HOTVR~\cite{Lapsien:2016zor} algorithm, as implemented in the
\FASTJET software package~\cite{Cacciari:2011ma}.
The anti-\kt jets are obtained using a distance parameter of $R=0.4$, and referred to as ``AK4 jets''.
For these jets, charged PF candidates are excluded from the clustering if their tracks are matched to
pileup vertices.
The HOTVR algorithm makes use of an effective distance parameter \Reff, scaling with $\rho/\pt$~\cite{Krohn:2009zg}.
The parameter $\rho$ controls the slope of \Reff and \pt denotes the jet \pt. The slope parameter is set to $\rho=600\GeV$.
The maximum of \Reff is set to 1.5, such that jets with $\pt<400\GeV$
do not have an active area~\cite{Cacciari:2007fd} larger than
Cambridge/Aachen~\cite{ca1,ca2} jets with $R=1.5$.
Jet grooming is performed during the jet clustering,
by suppressing the clustering of additional radiation into the jet using a veto based on
the mass jump algorithm~\cite{Stoll:2014hsa} in each iteration.
This veto rejects light clusters if the product of the combined mass of two clusters and
the mass jump parameter $\theta$ is smaller than the mass of the heavier cluster.
In this step, subjets are also identified.
The HOTVR parameters are set to the values described
in Ref.~\cite{Sirunyan:2020lcu}. The pileup per particle identification (PUPPI)
algorithm~\cite{Sirunyan:2020foa,Bertolini:2014bba}
is used to mitigate the effects of pileup, where HOTVR jets are clustered using PUPPI-corrected PF candidates.
Jet energy corrections~\cite{Khachatryan:2016kdb} are applied to AK4 jets and HOTVR subjets.
We have verified that the corrections derived for AK4 jets clustered with PUPPI
are suitable for correcting HOTVR subjets, using a sample enriched with events from \ttbar production.
The corrected HOTVR jet four-momentum is obtained from the sum of corrected subjet four-momenta.
The jet energy resolution in simulated events is smeared to match the resolution in data.

The presence of boosted \PQt quarks from the \bstar decay with Lorentz factors larger than approximately 1.5
allows one to reconstruct the full hadronic \PQt decay in a single HOTVR jet. Information about the jet substructure enables
discrimination of these \PQt jets from jets originating from light quarks and gluons.
A stable performance over a wide range of jet \pt is obtained by using HOTVR jets for \PQt tagging,
because of the adaptive jet size. The following selection criteria for HOTVR jets are used:
\begin{itemize}
\item the \pt fraction of the leading subjet $\mathrm{s}_1$ with respect to the jet, $f_{\pt} = \pt^{\mathrm{s}_1}/\pt^{\text{jet}} < 0.8$,
\item the number of subjets $N_{\text{sub}} \geq 3$,
\item the jet mass $140 < m_{\text{jet}} < 220 \GeV$, and
\item the minimum mass of pairs of subjets $m_{\text{min}} = \min(\sqrt{\smash{(P_i + P_j)}^2})>50\GeV$.
\end{itemize}
Additionally, a requirement on the ratio of the $N$-subjettiness~\cite{Thaler:2010tr,Thaler:2011gf} variables
$\tau_3/\tau_2 < 0.56$ is imposed to further increase the discrimination power against the QCD background~\cite{Sirunyan:2020lcu}.
The resulting \PQt tagging algorithm has an efficiency of 25\% at $\pt=200\GeV$, increasing to 40\% at
$\pt=2\TeV$, with a constant misidentification rate for jets from QCD multijet scattering of about 1\%.
A comprehensive comparison of \PQt tagging algorithms in CMS can be found in Ref.~\cite{Sirunyan:2020lcu}.
Covering a larger range in \PQt quark \pt than comparable tagging algorithms,
HOTVR offers a straightforward solution to extend the sensitivity of the
analysis from the highly boosted regime to a kinematic regime where the \PQt
quark decay products can be resolved, in a single approach.

The relatively long lifetime of bottom hadrons can result in a secondary vertex in events with
\PQb quarks. These secondary vertices can be reconstructed with information from the tracking detectors
in CMS and used to distinguish \PQb jets from light-quark and gluon jets (referred to in what follows as ``light jets").
The \textsc{DeepJet} \PQb tagging
algorithm~\cite{CMS-DP-2018-033} with the medium working point is used to identify AK4 \PQb jets arising from the \PQt quark decay,
with a selection efficiency of 80--85\% and a misidentification rate for light jets of 1\%.

The missing transverse momentum vector \ptvecmiss is computed as the negative vector sum of the transverse momenta
of all the PF candidates in an event, and its magnitude is denoted as \ptmiss~\cite{Sirunyan:2019kia}.
The \ptvecmiss is modified to account for corrections to the energy scale of the reconstructed AK4 jets in the event.

\section{Event selection}

In the offline analysis, we select events with exactly one isolated lepton with $\pt > 50\GeV$ and $\abs{\eta}<2.4$.
Events with an additional lepton with $\pt > 30\GeV$ and $\abs{\eta}<2.4$ are discarded.
The selected lepton does not overlap with any AK4 jet because of the requirement $\Delta R(\Pell, \text{jet}) > 0.4$.
To account for the presence of a neutrino from the leptonic \PW decay in the final state, $\ptmiss > 50\GeV$ is required and
\ptvecmiss must point in roughly the same azimuthal direction
(to within $\Delta\phi(\Pell, \ptvecmiss) < \pi/2$) as the momentum of the isolated lepton.
The SM backgrounds are suppressed by requiring $\HT > 200\GeV$, where \HT is the \pt-sum of all AK4 jets, as well as
$\ST>400\GeV$, where \ST is the sum of \HT, lepton \pt, and \ptmiss.
Finally, exactly one HOTVR jet with $\pt>200\GeV$ and $\abs{\eta}<2.5$ is required, fulfilling the HOTVR \PQt tagging criteria.

The \bstar mass is reconstructed from the mass of the $\PQt\PW$ system, \MtW.
The $\PQt\PW$ system is reconstructed from the \PQt-tagged HOTVR jet and the estimated \PW boson four-momentum, which is
obtained from the four-momentum of the charged lepton and \ptvecmiss.
Assuming that the \PW boson from the \bstar decay is on its mass shell, the neutrino is reconstructed
using the \PW boson mass as a constraint.
In cases where two solutions are found, the solution is selected that has the
smaller value of $\abs{p_{z,\PGn} - p_{z,\Pell}}$, where $p_{z,\PGn}$ and
$p_{z,\Pell}$ denote the components of the neutrino and charged lepton
four-momenta, respectively, along the beam direction.
The large boost of the \PW boson ensures that this solution gives an accurate reconstruction of the neutrino
four-momentum.

A $\chi^2$-like estimator \Xsquare is calculated for the reconstructed $\PQt\PW$ system, indicating how signal-like the event is,
under the assumption that the heavy resonance is produced at rest,
\begin{linenomath}
\begin{equation}
\Xsquare = \left(\frac{\Delta\phi_{\PQt,\PW} - \pi}{\sigma_{\Delta\phi_{\PQt,\PW}}}\right)^2
+ \left(\frac{A_{\pt}}{\sigma_{A_{\pt}}}\right)^2 .
\end{equation}
\end{linenomath}
The first term exploits the back-to-back signature of the signal. It becomes small if the azimuthal angular distance between the \PQt quark and the \PW boson candidates $\Delta\phi_{\PQt,\PW}$ is close to $\pi$.
The second term utilizes momentum conservation.
Since the \PQt quark and \PW boson originate from the decay of a heavy resonance in signal events,
their transverse momenta should be of similar size.
Hence, the second term becomes small if the \pt asymmetry
$A_{\pt}=(\pt^{\PQt} - \pt^{\PW})/(\pt^{\PQt} + \pt^{\PW})$
is close to zero.
The width parameters $\sigma_{\Delta\phi_{\PQt,\PW}}$ and $\sigma_{A_{\pt}}$ are determined from simulation.

Events passing the selection are categorized according to the number of \PQb-tagged AK4 jets.
Three categories are defined by having either zero (``0\PQb''), one (``1\PQb''), or more than one (``2\PQb'') \PQb-tagged
jets. The 0\PQb category is used to estimate the SM background containing no top quarks
(non-top background) from data, as described in Section~\ref{sec:background_estimation}.
The 2\PQb category is dominated by SM \ttbar production and is used as a control region to constrain
systematic uncertainties associated with the modeling of this background.
Two additional selection criteria are imposed in the 1\PQb category, to increase the signal sensitivity.
We require the \PQb-tagged jet to be at a large angular distance from the isolated lepton, $\Delta R(\Pell,\PQb) > 2.0$,
and the \Xsquare value must be smaller than 20.
Both requirements help to suppress the contribution from \ttbar production.
The 1\PQb category with these two additional requirements is the signal region of this search.
The signal efficiency for \bstar quarks decaying in the \ljets final state is between 4 and 9\% for the LH model, and 4 to 10\% for the RH model.
The efficiency for the RH model is slightly higher, because of the harder jet \pt spectra.

\section{Background estimation}\label{sec:background_estimation}

The dominant background in the selected phase space of the signal region originates from
SM production of top quarks, \ttbar and single \PQt, and is constrained using the 2\PQb control region.
Top tagging scale factors~\cite{Sirunyan:2020lcu} are applied to the simulated samples,
in order to correct for differences in \PQt tagging efficiencies.
The scale factors are found to be consistent with unity.

Contributions from events with a misidentified \PQt jet are estimated from data in the 0\PQb category.
By requiring a \PQt-tagged HOTVR jet, but vetoing \PQb-tagged AK4 jets,
this category consists dominantly of events from \WZjets and diboson production with misidentified \PQt jets.
The $\alpha$-ratio method~\cite{Sirunyan:2016cao} is used to extrapolate the normalization and shape in \MtW of these backgrounds
into the 1\PQb signal and 2\PQb control regions.
The ratio $\alpha$ is defined as the number of events in the 1\PQb or 2\PQb category to the number of events in the 0\PQb
category for a given bin in \MtW. It is calculated from simulated \WZjets and diboson samples. By multiplying the observed
number of events in data in the 0\PQb category with the ratio $\alpha$, the background prediction for the 1\PQb and 2\PQb
categories is obtained.
A fit to the distributions of $\alpha$ is performed to obtain a smooth background prediction with reduced sensitivity to
bin-by-bin statistical fluctuations in the simulated samples. The systematic uncertainty arising from the choice of fit function
is estimated using different fit parameterizations and is added in
quadrature to the statistical uncertainty.

The resulting functions $\alpha(\MtW)$ are used to estimate the non-top backgrounds in the 1\PQb and 2\PQb categories.
In order to keep the background estimation in the 1\PQb and 2\PQb categories statistically independent,
we split the data in the 0\PQb category randomly into two subsets, thus two thirds of the data are used to estimate the
non-top background in the 1\PQb category and one third of the data are used in the 2\PQb category.
The statistical uncertainty from the non-top background obtained from data in the 0\PQb category is negligible
compared to other uncertainties, such that the exact proportions of the subsets are not important.

\section{Systematic uncertainties}\label{sec:sys_unc}

\begin{table}[tb]
\centering
\topcaption{Summary of all considered sources of systematic uncertainties affecting the \MtW
distributions in the 1\PQb and 2\PQb categories of the \ljets channel.
The source of the uncertainty is given in the first column.
The second column indicates if the uncertainty results in a change of normalization or shape of the
\MtW distribution. The samples affected by a given uncertainty source are shown in the third column.
The fourth column shows the impact of these uncertainties, estimated for an LH \bstar signal with a mass
of 2.4\TeV. These are quantified by calculating the change in the fitted signal strength when a
given parameter is displaced by $\pm1$ standard deviation from its post-fit value, divided by the total uncertainty in the fitted signal.
Uncertainties taken to be fully correlated across the
three years are given in the upper part of the table.
Uncertainties affecting both the \ljets and all-hadronic channels are marked by an asterisk.
\label{tab:syst_uncertainties}}

\begin{tabular}{ l c l c }
Source & Uncertainty & Samples & Impact (up/down) \Bstrut \\ \hline
\ttbar cross section$^\ast$ & $\pm$ 20\% & \ttbar & $+5.1$ / $-5.0$\% \Tstrut \\
Single \PQt cross section$^\ast$ & $\pm$ 30\% & single \PQt & $-4.2$ / $+4.8$\% \\
Luminosity$^\ast$ & $\pm$ 1.6\% & \ttbar, single \PQt, signal & $-1.4$ / $+1.5$\% \\
Top quark \pt reweighting$^\ast$ & Shape & \ttbar & $-5.0$ / $+5.1$\% \\
PDF$^\ast$ & Shape & \ttbar, single \PQt, signal & $-4.1$ / $+4.6$\% \\
Background estimation (1\PQb) & Shape & non-top (from data) & $-5.3$ / $+6.9$\% \\
Background estimation (2\PQb)& Shape & non-top (from data) & $-0.3$ / $-0.4$\% \Bstrut \\ \hline
Pileup$^\ast$ & Shape & \ttbar, single \PQt, signal & $-0.4$ / $+0.5$\% \Tstrut \\
JES$^\ast$ & Shape & \ttbar, single \PQt, signal & $-1.3$ / $+2.3$\% \\
JER$^\ast$ & Shape & \ttbar, single \PQt, signal & $+0.0$ / $+0.4$\% \\
ECAL trigger timing$^\ast$ & Shape & \ttbar, single \PQt, signal & $+0.1$ / $-0.0$\% \\
Electron identification & Shape & \ttbar, single \PQt, signal & $+0.4$ / $-0.4$\% \\
Electron reconstruction & Shape & \ttbar, single \PQt, signal & $+0.2$ / $-0.0$\% \\
Electron trigger & Shape & \ttbar, single \PQt, signal & $+0.3$ / $-0.0$\% \\
Muon identification & Shape & \ttbar, single \PQt, signal & $+0.4$ / $-0.1$\% \\
Muon isolation & Shape & \ttbar, single \PQt, signal & $+0.3$ / $-0.1$\% \\
Muon trigger & Shape & \ttbar, single \PQt, signal & $-0.1$ / $+0.4$\% \\
\PQt tagging (fully merged) & Shape & \ttbar, single \PQt, signal & $-1.2$ / $+1.5$\% \\
\PQt tagging (partially merged) & Shape & \ttbar, single \PQt, signal & $-0.7$ / $+0.8$\% \\
\PQt tagging (nonmerged) & Shape & \ttbar, single \PQt, signal & $-0.0$ / $+0.2$\% \\
\PQb tagging (\PQb, \cPqc) & Shape & \ttbar, single \PQt, signal & $-3.6$ / $+4.0$\% \\
\PQb tagging (\cPqu, \cPqd, \cPqs, \cPg) & Shape & \ttbar, single \PQt, signal & $+0.7$ / $-0.6$\% \\
\end{tabular}

\end{table}

Several systematic uncertainties are taken into account in the analysis,
affecting the normalization and shape of the final \MtW distributions in the 1\PQb and 2\PQb categories.
A summary of all considered sources is given in Table~\ref{tab:syst_uncertainties}.
Some of these uncertainties are assumed to be fully correlated throughout all three years;
these uncertainties are given in the upper part of the table.
The contributions of uncertainties without year-to-year correlations
are calculated for each year independently.

In the following, a detailed description of the considered sources of systematic uncertainties is given.
\begin{itemize}
\item The total integrated luminosity of the data set is assigned an
uncertainty of 1.6\%~\cite{CMS:2021xjt,CMS:2018elu,CMS:2019jhq}.
\item We assign uncertainties of 20 and 30\% to the \ttbar and single top quark production
cross sections, respectively. These account for uncertainties due to missing higher orders, estimated by
halving and doubling the renormalization and factorization scales in the corresponding simulations,
and for the uncertainties from the normalization to the NNLO and NLO predictions.
\item The uncertainty from the choice of PDFs is estimated by calculating the signal and background predictions in each
bin of the \MtW distribution using 100 replicas of the NNPDF sets~\cite{Butterworth:2015oua}.
The standard deviation of these predictions is used to construct shape variations.
In the case of the signal samples, the distributions are normalized to the respective cross sections,
such that only the acceptance effects are considered.
\item The uncertainty in the pileup distribution is estimated by varying the total inelastic cross section
of 69.2\unit{mb} within the assigned uncertainty of $\pm5$\%~\cite{Sirunyan:2018nqx},
and propagating the changes to the event yields in the simulation.
\item The jet energy scale (JES) and jet energy resolution (JER) corrections are varied within their uncertainties to
derive their effect on the shapes of the \MtW distributions.
\item A timing shift of the signals in ECAL cells with respect to the L1 trigger
clock led to a trigger inefficiency in the years 2016 and 2017 for events with
significant amount of energy in the ECAL in the region $2.0 < \eta < 3.0$.
The simulations were corrected for this effect. The associated systematic
uncertainties are estimated by varying the corrections within their
uncertainties.
\item Uncertainties from the electron and muon identification and reconstruction are estimated by
varying the corresponding efficiency corrections within their uncertainties.
\item Trigger efficiency corrections are also varied within their uncertainties.
\item The transfer functions $\alpha$ used for the background estimations of the non-top backgrounds
in the 1\PQb and 2\PQb categories are varied within their uncertainties, as discussed above.
\item Differences in the \PQt and \PQb tagging efficiencies between data and simulation are accounted for
by data-to-simulation scale factors. These scale factors are varied within their uncertainties.
The HOTVR \PQt tagging scale factors are split into fully merged, partially merged, and nonmerged cases,
depending on how many partons from the \PQt decay can be matched to the HOTVR jet~\cite{Sirunyan:2020lcu}. These scale factors are
varied independently to obtain the effect on the \MtW distributions.
The \PQb tagging scale factors are split into efficiencies for \PQb, \cPqc, and light jets~\cite{Sirunyan:2017ezt}.
The uncertainties for \PQb and \cPqc jets are taken to be correlated and varied simultaneously,
while the uncertainties for light jets are taken to be uncorrelated and varied independently.
\item The uncertainty arising from the modeling of the top quark \pt spectrum is estimated by varying the parameter $\beta$,
affecting the shape of the \pt distribution, by $\pm$50\%. This allows the fit to determine the correct shape of the \ttbar background
and constrain the associated uncertainty from the 2\PQb control region.
\end{itemize}

The results of this analysis are combined with a previously
published analysis in the all-hadronic final state~\cite{Sirunyan:2021fkj}.
Systematic uncertainties affecting both final states are taken to be fully correlated
between the two analyses, and are indicated in Table~\ref{tab:syst_uncertainties} by
an asterisk. Systematic uncertainties affecting only the all-hadronic channel are
not shown, but are described in detail in Ref.~\cite{Sirunyan:2021fkj}.

\section{Results\label{sec:results}}

\begin{figure}[tb]
\centering
\includegraphics[width=.48\textwidth]{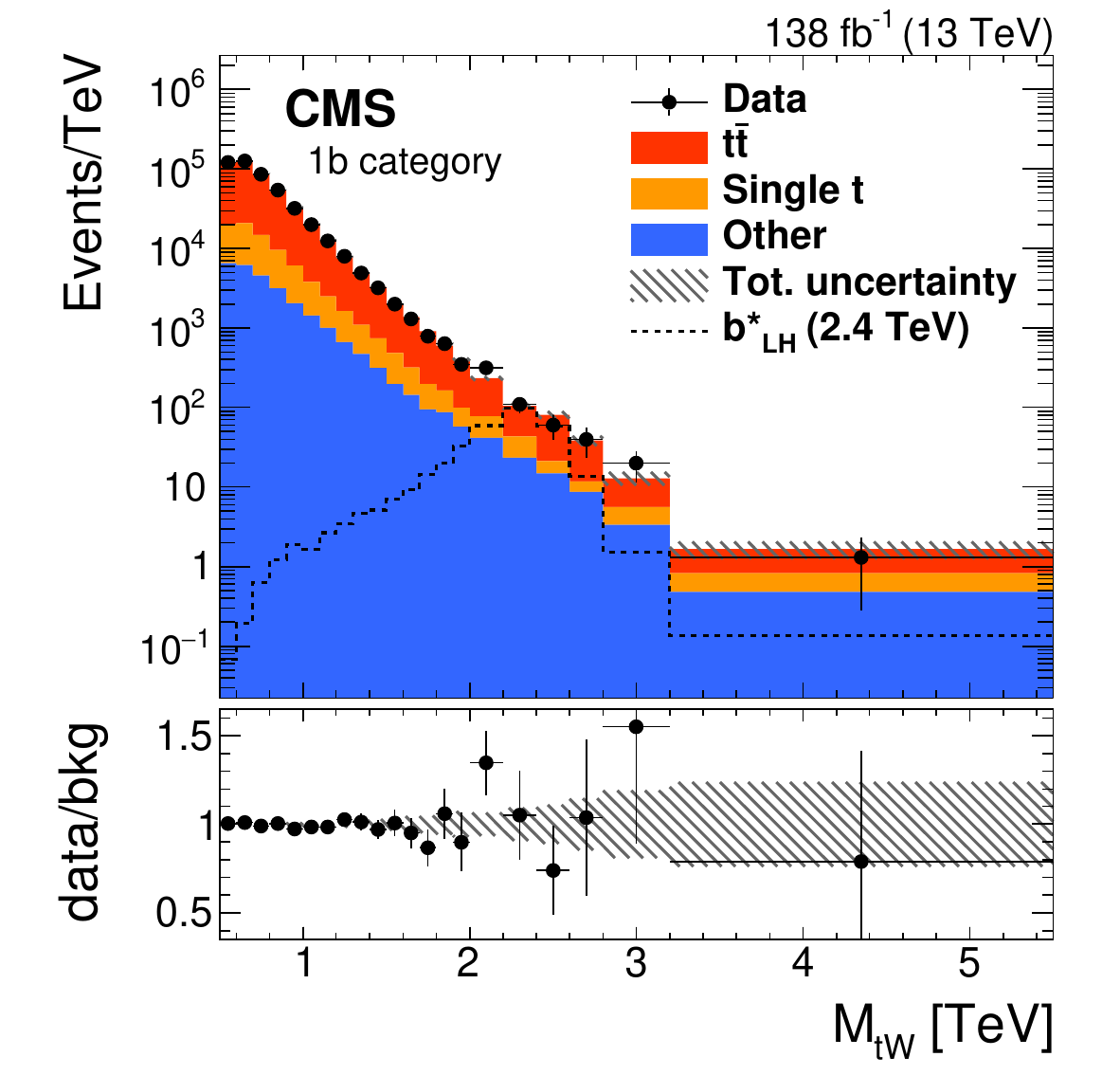}
\includegraphics[width=.48\textwidth]{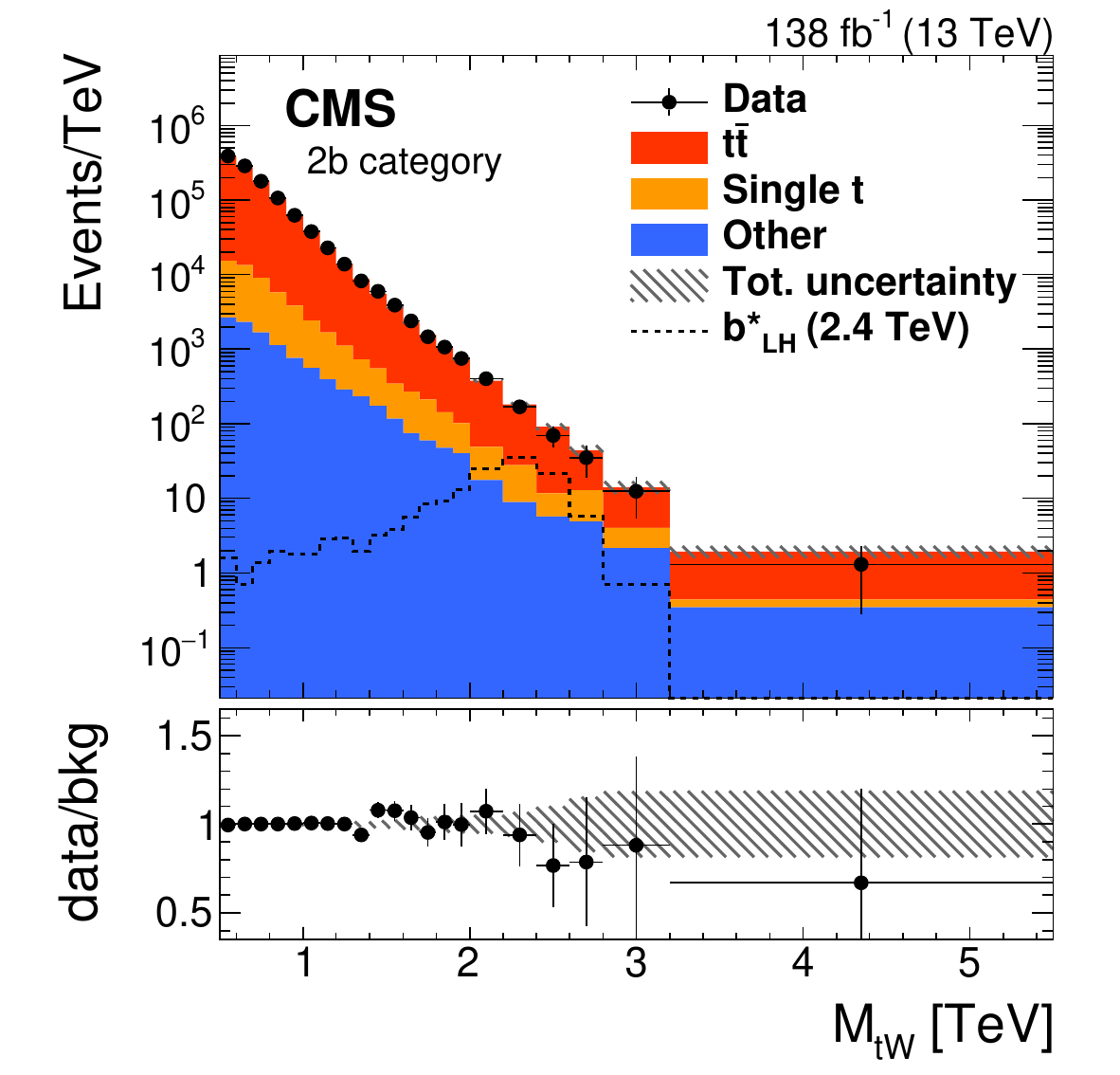}
\caption{Distributions of \MtW in the 1\PQb (left) and 2\PQb (right) categories.
The data are shown by filled markers, where the horizontal bars indicate the bin widths.
The individual background contributions are given by filled histograms.
The expected signal for an LH \bstar with a mass of $m_{\bstar}=2.4\TeV$ is shown by a dashed line.
The shaded region is the uncertainty in the total background estimate.
The lower panels of each figure show the ratio of data to the background estimate,
with the total uncertainty in the predicted background displayed as shaded band.
\label{fig:postfit_distribution}}
\end{figure}

For the statistical interpretation of the results, a combination with the
analysis in the all-hadronic final state~\cite{Sirunyan:2021fkj} is performed.
That analysis considers events with a dijet topology, with two highly energetic
jets with a large azimuthal separation. One jet is required to be \PQt tagged, the other
one is \PW tagged. The dominant background from multijet production with two misidentified
jets is obtained from a control region with inverted \PQt tagging requirements.
The expected multijet background is extrapolated to the signal region with the help
of a two-dimensional pass-fail-ratio in the jet mass of the top-tagged jet and \MtW.
An additional control region enriched by events from \ttbar production is considered
in order to constrain this background.
Events containing isolated leptons with $\pt>30\GeV$ are rejected in the selection of the all-hadronic
regions, to avoid double counting events that also appear in the \ljets regions.

\begin{figure}[pt]
\centering
\includegraphics[width=.48\textwidth]{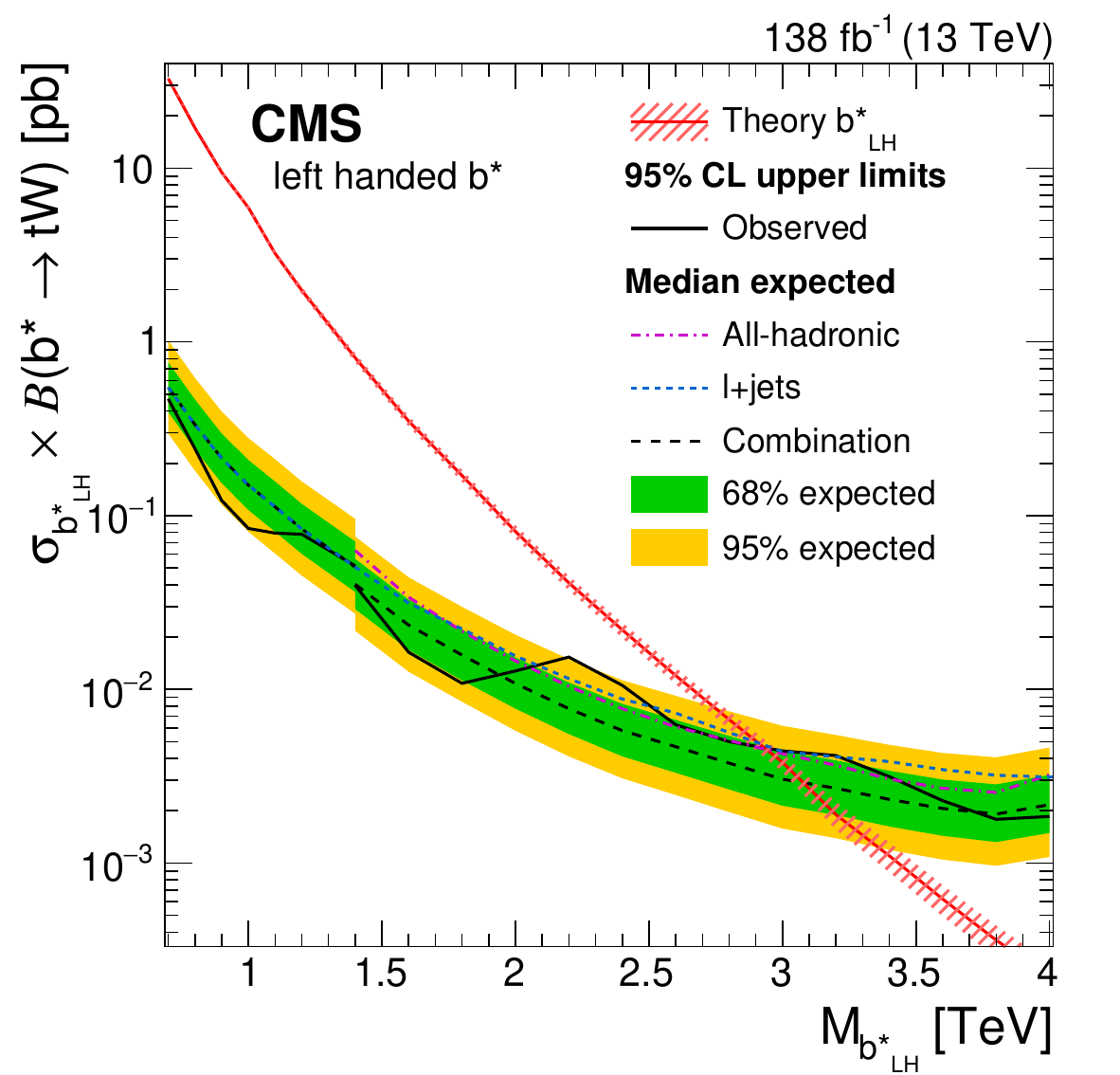}
\includegraphics[width=.48\textwidth]{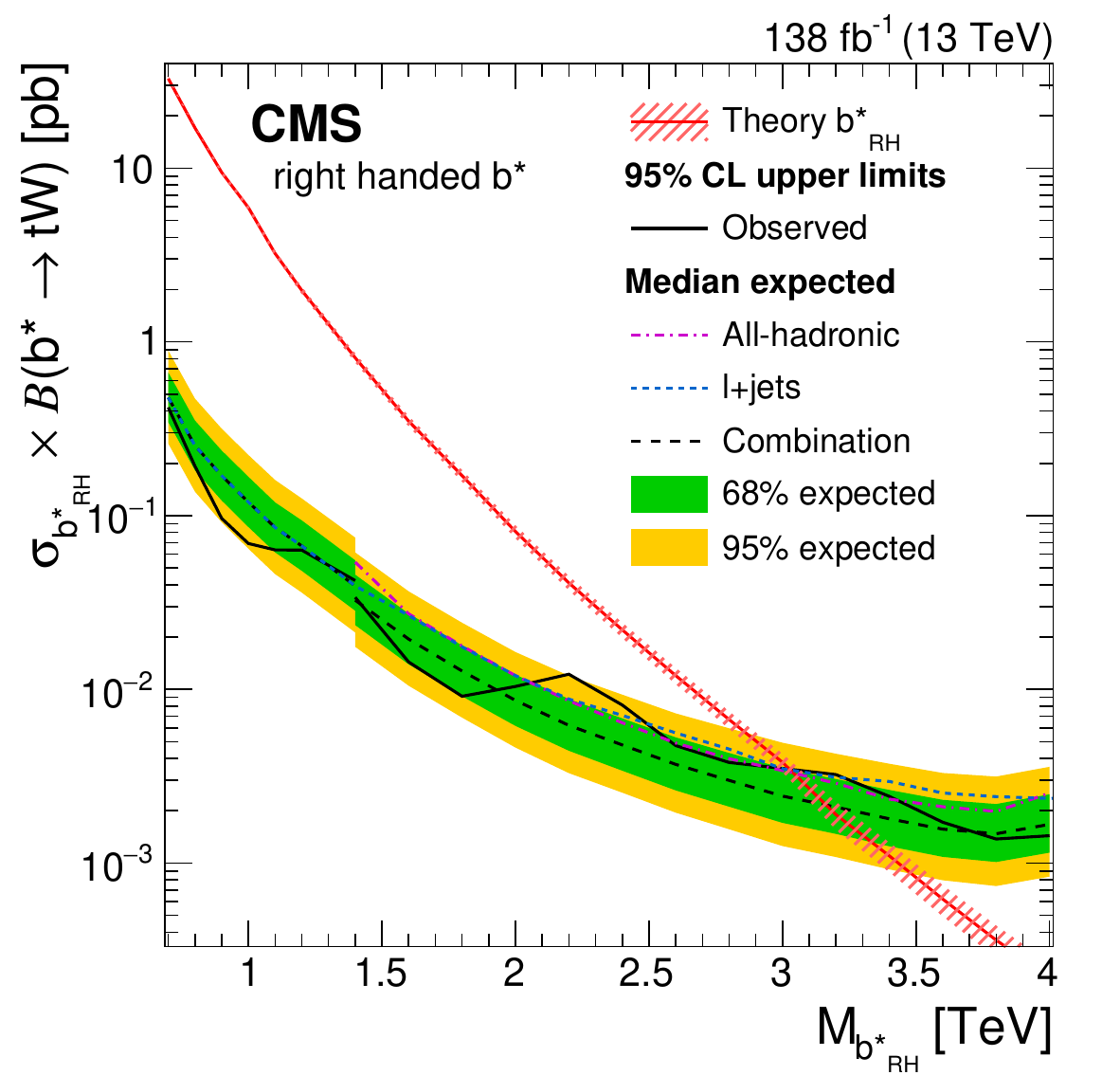}\\
\includegraphics[width=.48\textwidth]{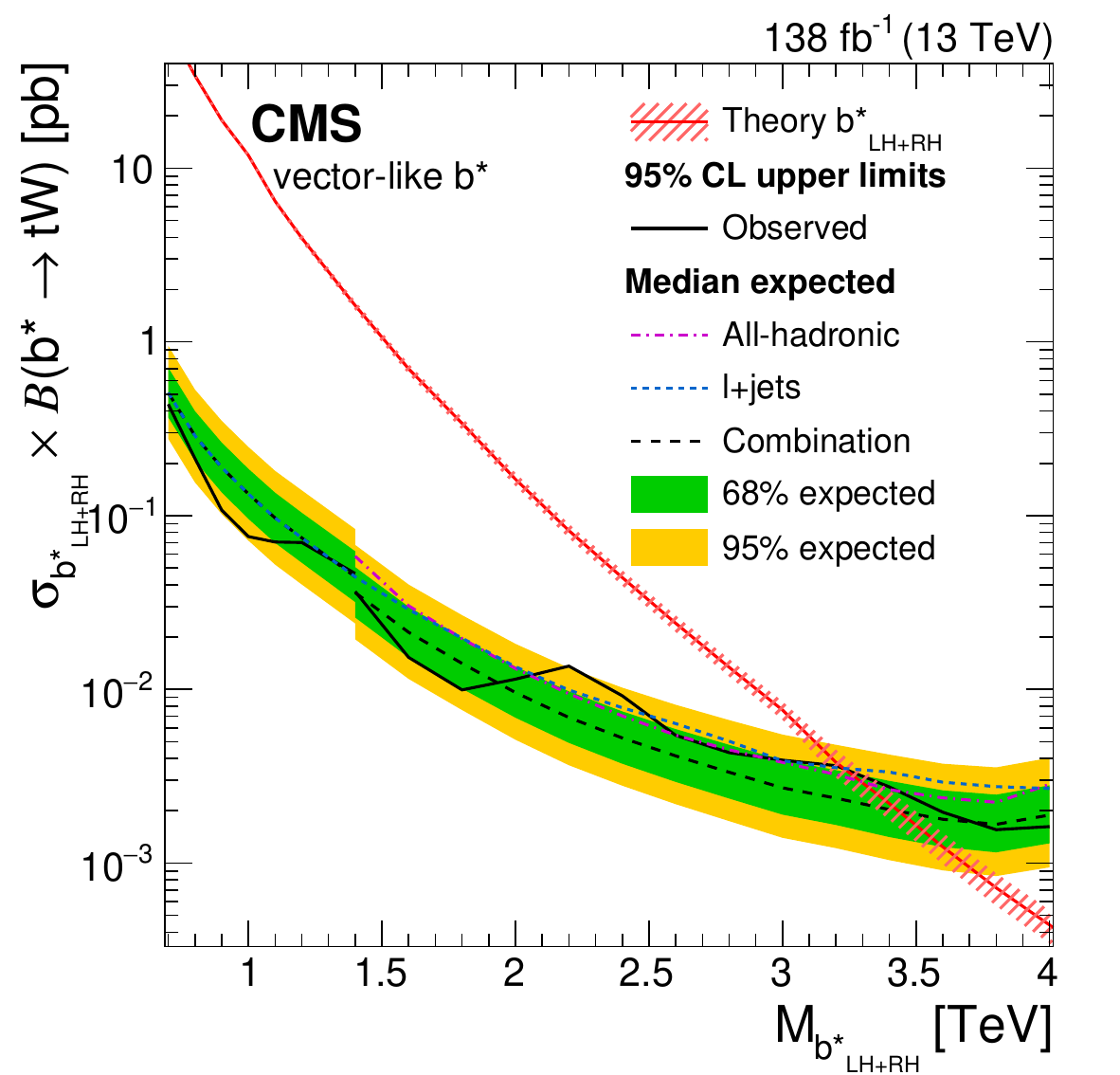}
\caption{Upper limits on the product of production cross section and branching fraction of the left-handed (upper left), right-handed (upper right)
and vector-like (lower) \bstar hypotheses at 95\% \CL.
Colored lines show the expected limits from the \ljets (dotted) and all-hadronic (dash-dotted) channels,
where the latter start at \bstar masses of 1.4\TeV.
The observed and expected limits from the combination are shown as solid and dashed black lines, respectively.
The green and yellow bands show the 68 and 95\% confidence intervals on the combined expected limits.
The theoretical cross sections are shown as the red lines, where the uncertainties due to missing higher orders
are depicted by shaded areas.
\label{fig:bstar_limits}}
\end{figure}

A fit is performed simultaneously to the \MtW distributions, not only in the 1\PQb and 2\PQb categories of this analysis,
but also to the signal and multijet control regions in the earlier all-hadronic analysis.
The \ttbar control region of the all-hadronic
analysis is not considered, because of the higher precision of the 2\PQb category.
The contributions from all background and signal processes are fitted
simultaneously to data in a binned maximum likelihood fit.
The signal strength is a free parameter in the model and systematic uncertainties are accounted for
by nuisance parameters assigned to the sources of systematic uncertainties described in Section~\ref{sec:sys_unc}.
These are profiled in the fit.
Normalization uncertainties are modeled with log-normal priors, and uncertainties affecting
shapes are modeled using a template morphing approach with Gaussian priors~\cite{Conway:2011in}.
Statistical uncertainties are treated in a bin-by-bin approach using a simplified version of
the Barlow--Beeston method~\cite{Barlow:1993dm}.
The post-fit distributions of \MtW in the 1\PQb and 2\PQb categories are shown in Fig.~\ref{fig:postfit_distribution}.
Data and background predictions are found to be statistically in agreement over the full spectrum of \MtW.

No significant excess of data over the expected SM background is observed. Upper limits on
the product of production cross section and branching fraction into \PQt{}\PW of the three coupling scenarios in the
\bstar benchmark model at 95\% \CL are set using the \CLs method~\cite{Junk:1999kv,Read:2002hq}
with an asymptotic approximation to the profile likelihood test statistic~\cite{Cowan:2010js}.
Figure~\ref{fig:bstar_limits} shows the expected and observed
limits of the combination, as well as the expected limits of the two separate analyses in the all-hadronic and \ljets final states.
The step at $\MtW=1.4\TeV$ in the combined limits arises from the extended reach of the \ljets channel
towards lower masses, where it places unique constraints in the range 0.7--1.4\TeV. The sensitivity for masses
above 1.4\TeV is comparable between the all-hadronic and \ljets channels, resulting in stricter
limits compared to the separate analyses.
We derive \bstar mass exclusion limits by comparing the upper cross section limits with
theoretical cross sections of \bstar production, calculated with \MGvATNLO.
The observed (expected) mass limits are 3.0, 3.0, and 3.2\TeV (3.1, 3.2, and 3.4\TeV) for the LH,
RH, and VL hypotheses, respectively. These are the most stringent constraints on this model
to date.

\begin{figure}[tb]
\centering
\includegraphics[width=.48\textwidth]{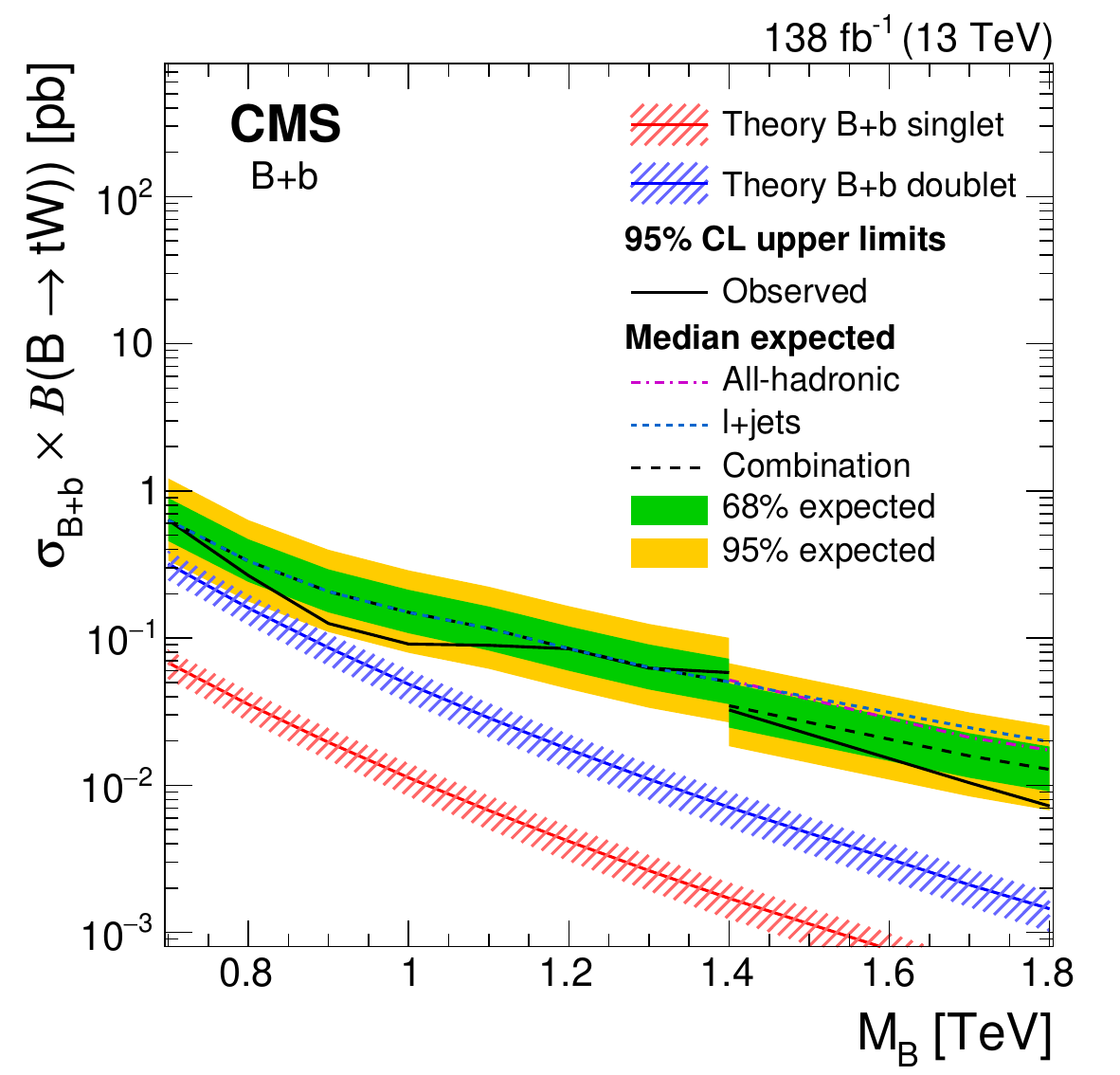}
\includegraphics[width=.48\textwidth]{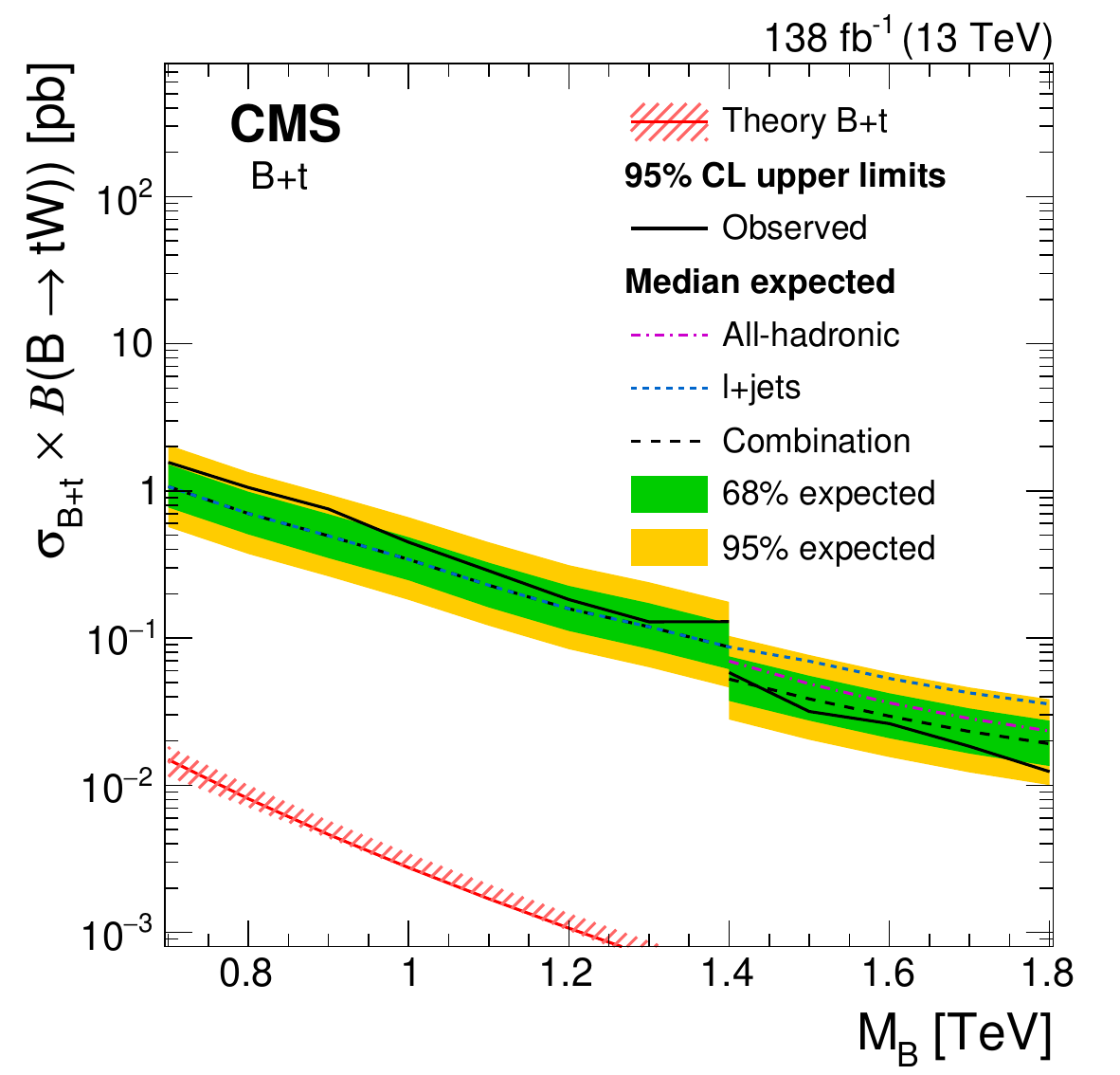}
\caption{Upper limits on the product of production cross section and branching fraction of
the \PBb (left) and \PBt (right) production modes at 95\% \CL.
Colored lines show the expected limits from the \ljets (dotted) and all-hadronic (dash-dotted) channels,
where the latter start at \PB masses of 1.4\TeV.
The observed and expected limits from the combination are shown as solid and dashed black lines, respectively.
The green and yellow bands show the 68 and 95\% confidence intervals on the combined expected limits.
The theoretical cross sections are shown as the red and blue lines, where the uncertainties due to missing higher orders
are depicted by shaded areas.
\label{fig:bprime_limits}}
\end{figure}

The results can also be interpreted in the context of a singly produced vector-like \PB quark decaying into \PQt{}\PW.
The production modes \PBt and \PBb are tested separately and the upper limits on the product of production cross sections and
branching fraction at 95\% \CL are shown in Fig.~\ref{fig:bprime_limits}. Over a \PB mass range from 0.7 to 1.8\TeV, the observed
upper limits range from 0.63 to 0.007\unit{pb} for the \PBb production mode, and from 1.6 to 0.01\unit{pb} for \PBt production.
The upper cross section limits are compared to the theory predictions from Ref.~\cite{Carvalho:2018jkq}.
Models with a VLQ singlet or VLQ doublet are considered in case of \PBb production.
For \PBt production, both models result in the same production cross section.
Because of the small theoretical cross sections for these processes, no mass limits are set.

The upper cross section limits can be compared to a previous search for singly produced vector-like B quarks at $\sqrt{s}=13\TeV$
using a data set corresponding to 35.9\fbinv by the CMS collaboration~\cite{CMS:2018dcw}.
For the \PBb production mode, the expected limits are 8--29\% better at low \PB masses,
where the \ljets channel alone sets limits, and 75\% better at the highest considered \PB mass.
For the \PBt production mode, the \ljets channel is less sensitive,
because of the additional \PQb jet from the decay of the associated \PQt quark, leading to weaker limits at low \PB masses
compared to the previous search.
At high \PB masses, this search improves the upper limits by up to 50\%.

\section{Summary}

A search for a heavy resonance decaying to $\PQt\PW$ in the final state with a
lepton and a \PQt-tagged jet has been presented.
The data analyzed correspond to an integrated luminosity of 138\fbinv of proton-proton
collisions collected at a center-of-mass energy of 13\TeV.
Final states where the \PW boson decays leptonically and the top quark decay results in a single jet are probed.
Compared to an earlier analysis of the all-hadronic final state, the lower reach of the analysis is extended from 1.4 down to 0.7\TeV, because of lower lepton trigger thresholds and the extended range in \PQt quark transverse momentum provided by the Heavy Object Tagger with Variable R.
Above 1.4\TeV a combination with the search in the all-hadronic final state has been performed.
The dominant \ttbar background is constrained using a dedicated control region
and the background from misidentified \PQt jets is estimated from data.
No significant excess of data over the background prediction is observed.
Upper limits on the single production of a vector-like quark decaying to $\PQt\PW$ have been derived
in the mass range of 0.7 to 1.8\TeV.
The excited bottom quark \bstar hypotheses with left-handed, right-handed, and vector-like chiralities are excluded at 95\% confidence level
up to masses of 3.0, 3.0, and 3.2\TeV, respectively. The upper limits on the product of
cross section and branching fraction represent the most stringent constraints on the \bstar model to date.

\begin{acknowledgments}
    We congratulate our colleagues in the CERN accelerator departments for the excellent performance of the LHC and thank the technical and administrative staffs at CERN and at other CMS institutes for their contributions to the success of the CMS effort. In addition, we gratefully acknowledge the computing centers and personnel of the Worldwide LHC Computing Grid and other centers for delivering so effectively the computing infrastructure essential to our analyses. Finally, we acknowledge the enduring support for the construction and operation of the LHC, the CMS detector, and the supporting computing infrastructure provided by the following funding agencies: BMBWF and FWF (Austria); FNRS and FWO (Belgium); CNPq, CAPES, FAPERJ, FAPERGS, and FAPESP (Brazil); MES and BNSF (Bulgaria); CERN; CAS, MoST, and NSFC (China); MINCIENCIAS (Colombia); MSES and CSF (Croatia); RIF (Cyprus); SENESCYT (Ecuador); MoER, ERC PUT and ERDF (Estonia); Academy of Finland, MEC, and HIP (Finland); CEA and CNRS/IN2P3 (France); BMBF, DFG, and HGF (Germany); GSRI (Greece); NKFIA (Hungary); DAE and DST (India); IPM (Iran); SFI (Ireland); INFN (Italy); MSIP and NRF (Republic of Korea); MES (Latvia); LAS (Lithuania); MOE and UM (Malaysia); BUAP, CINVESTAV, CONACYT, LNS, SEP, and UASLP-FAI (Mexico); MOS (Montenegro); MBIE (New Zealand); PAEC (Pakistan); MSHE and NSC (Poland); FCT (Portugal); JINR (Dubna); MON, RosAtom, RAS, RFBR, and NRC KI (Russia); MESTD (Serbia); SEIDI, CPAN, PCTI, and FEDER (Spain); MOSTR (Sri Lanka); Swiss Funding Agencies (Switzerland); MST (Taipei); ThEPCenter, IPST, STAR, and NSTDA (Thailand); TUBITAK and TAEK (Turkey); NASU (Ukraine); STFC (United Kingdom); DOE and NSF (USA).
     
    \hyphenation{Rachada-pisek} Individuals have received support from the Marie-Curie program and the European Research Council and Horizon 2020 Grant, contract Nos.\ 675440, 724704, 752730, 758316, 765710, 824093, 884104, and COST Action CA16108 (European Union); the Leventis Foundation; the Alfred P.\ Sloan Foundation; the Alexander von Humboldt Foundation; the Belgian Federal Science Policy Office; the Fonds pour la Formation \`a la Recherche dans l'Industrie et dans l'Agriculture (FRIA-Belgium); the Agentschap voor Innovatie door Wetenschap en Technologie (IWT-Belgium); the F.R.S.-FNRS and FWO (Belgium) under the ``Excellence of Science -- EOS" -- be.h project n.\ 30820817; the Beijing Municipal Science \& Technology Commission, No. Z191100007219010; the Ministry of Education, Youth and Sports (MEYS) of the Czech Republic; the Deutsche Forschungsgemeinschaft (DFG), under Germany's Excellence Strategy -- EXC 2121 ``Quantum Universe" -- 390833306, and under project number 400140256 - GRK2497; the Lend\"ulet (``Momentum") Program and the J\'anos Bolyai Research Scholarship of the Hungarian Academy of Sciences, the New National Excellence Program \'UNKP, the NKFIA research grants 123842, 123959, 124845, 124850, 125105, 128713, 128786, and 129058 (Hungary); the Council of Science and Industrial Research, India; the Latvian Council of Science; the Ministry of Science and Higher Education and the National Science Center, contracts Opus 2014/15/B/ST2/03998 and 2015/19/B/ST2/02861 (Poland); the Funda\c{c}\~ao para a Ci\^encia e a Tecnologia, grant CEECIND/01334/2018 (Portugal); the National Priorities Research Program by Qatar National Research Fund; the Ministry of Science and Higher Education, projects no. 14.W03.31.0026 and no. FSWW-2020-0008, and the Russian Foundation for Basic Research, project No.19-42-703014 (Russia); the Programa Estatal de Fomento de la Investigaci{\'o}n Cient{\'i}fica y T{\'e}cnica de Excelencia Mar\'{\i}a de Maeztu, grant MDM-2015-0509 and the Programa Severo Ochoa del Principado de Asturias; the Stavros Niarchos Foundation (Greece); the Rachadapisek Sompot Fund for Postdoctoral Fellowship, Chulalongkorn University and the Chulalongkorn Academic into Its 2nd Century Project Advancement Project (Thailand); the Kavli Foundation; the Nvidia Corporation; the SuperMicro Corporation; the Welch Foundation, contract C-1845; and the Weston Havens Foundation (USA).
\end{acknowledgments}

\bibliography{auto_generated}
\cleardoublepage \appendix\section{The CMS Collaboration \label{app:collab}}\begin{sloppypar}\hyphenpenalty=5000\widowpenalty=500\clubpenalty=5000\cmsinstitute{Yerevan~Physics~Institute, Yerevan, Armenia}
A.~Tumasyan
\cmsinstitute{Institut~f\"{u}r~Hochenergiephysik, Vienna, Austria}
W.~Adam\cmsorcid{0000-0001-9099-4341}, J.W.~Andrejkovic, T.~Bergauer\cmsorcid{0000-0002-5786-0293}, S.~Chatterjee\cmsorcid{0000-0003-2660-0349}, K.~Damanakis, M.~Dragicevic\cmsorcid{0000-0003-1967-6783}, A.~Escalante~Del~Valle\cmsorcid{0000-0002-9702-6359}, R.~Fr\"{u}hwirth\cmsAuthorMark{1}, M.~Jeitler\cmsAuthorMark{1}\cmsorcid{0000-0002-5141-9560}, N.~Krammer, L.~Lechner\cmsorcid{0000-0002-3065-1141}, D.~Liko, I.~Mikulec, P.~Paulitsch, F.M.~Pitters, J.~Schieck\cmsAuthorMark{1}\cmsorcid{0000-0002-1058-8093}, R.~Sch\"{o}fbeck\cmsorcid{0000-0002-2332-8784}, D.~Schwarz, S.~Templ\cmsorcid{0000-0003-3137-5692}, W.~Waltenberger\cmsorcid{0000-0002-6215-7228}, C.-E.~Wulz\cmsAuthorMark{1}\cmsorcid{0000-0001-9226-5812}
% ...
\vskip\cmsinstskip
% ...
1:~Also at TU Wien, Wien, Austria\\
% ...
\end{sloppypar}
\end{document}